\documentstyle[aps,preprint]{revtex}
\tightenlines
\begin{document}
\begin{center}
{\Large{\bf Temperature induced shell effects in deformed nuclei}}\\
B. K. Agrawal$^1$, Tapas Sil$^1$, S. K. Samaddar$^1$ and  J. N. De$^{2}$ \\
$^1$Saha Institute of Nuclear Physics, 1/AF Bidhannagar, Calcutta 700 064,
India\\
$^2$Variable Energy Cyclotron Centre, 1/AF Bidhannagar, Calcutta 700 064,
India\\
\end{center}
\begin{abstract}
The thermal evolution of the shell correction energy is investigated for
deformed nuclei using Strutinsky prescription in a self-consistent
relativistic mean-field framework. For temperature independent
single-particle states corresponding to either spherical or deformed
nuclear shapes, the shell correction energy $\Delta_{sc}$ steadily
washes out with temperature. However, for states pertaining to
the self-consistent thermally evolving shapes of deformed nuclei,
the dual role played by the single-particle occupancies in diluting
the fluctuation effects from the single-particle spectra and in 
driving the system towards a smaller deformation is crucial in determining
$\Delta_{sc}$ at moderate temperatures. In rare earth nuclei, it is 
found that $\Delta_{sc}$ builds up strongly around the shape transition
temperature; for lighter deformed nuclei like $^{64}Zn$
and $^{66}Zn$, this is relatively less prominent.
\end{abstract}

\vskip 0.5cm
PACS number(s): 21.10.Ma, 21.60.-n, 27.70.+q
\newpage

In the backdrop of the nuclear mass formula, the shell-correction energy
calculated in the microscopic-macroscopic approach of Strutinsky \cite{str}
plays an extremely important
role in properly understanding the ground state masses
of atomic nuclei, their equilibrium deformation, double-hump barrier of
fissioning nuclei etc. The shell correction energy is evaluated through
smoothening of the fluctuations in the single-particle level  structure;
these fluctuations depend among others on the intrinsic deformation of the
nuclear system. With increasing excitation or temperature, occupancies of
orbitals tend to wash away the fluctuations arising from the set of
single-particle states; it is then found that the shell energy melts away
\cite{ram} with temperature at $T_{shell} \sim \hbar\omega_0/\pi$
($\hbar\omega_0 \sim 41 A^{-1/3}$). This is typically $\sim 2.5 $ MeV for
rare-earth nuclei. This has however been investigated for a fixed set of
single-particle levels corresponding to a fixed intrinsic
deformation. 

Nuclei, deformed in their ground states undergo phase transition to
spherical shapes with increase in temperature \cite{bra,goo}. For
rare-earth nuclei, these temperatures $T_p$ are typically between  $1 - 2$
MeV. In their ground states, the $m-$states emanating from a given
$j-$shell near the Fermi surface of these nuclei are preferentially  occupied
driving the system towards a stable deformation. As the nuclei are heated
up, these occupancies tend to equalize which tries to restore the
spherical symmetry. The different $m-$states from a given orbital then
become bunched and the single-particle level structure becomes more
nonuniform at the spherical shape ($\beta_2 = 0$). If  one calculates the
shell correction energy at $T =0$ with the deformed ground state level
spectrum and the spectrum  obtained by constraining the system at $\beta_2
=0$, one would always find that  the shell correction energy at $\beta_2 =0$
would be larger in magnitude. With increasing temperature, the
deformation of the nuclei reduces and the occupancies of the
single-particle states in the vicinity of the Fermi surface tend to
become uniform. The uniform occupancy across the Fermi surface 
would reduce the shell correction whereas the decrease of deformation 
enhances the shell correction. In a self-consistent tuning-in of the
deformation with temperature, the temperature dependence  of the shell
correction energy would then be governed by the dual role played by the
thermal evolution of the single-particle occupancies, i.e., the interplay
between their tendency for a drive towards sphericity and their tendency
for the washing out of the shell corrections. For example, if $T_p$ is 
considerably less than $T_{shell}$, 
one may  find a sharp building-up of the shell
correction energy with temperature in the vicinity of $T_p$. The aim of
this note is to investigate in detail  this delicate interplay for a few selected nuclei in
the relativistic mean field (RMF) theory. For our study, we have chosen 
isotopes of $Sm$ and $Dy$ in the rare-earth region and two lighter nuclei 
$^{64}Zn$ and $^{66}Zn$.
In passing, it may be mentioned that Egido {\it et al} \cite{egi} have
recently pointed out that at relatively high temperature, the shell-effects
which drive deformation disappear but not the ones providing magic numbers
in spherical nuclei.

The details of the Lagrangian density and the corresponding field
equations used for calculating the thermal
evolution of the deformation and single-particle levels are given in Refs.
\cite{gam,agr}. In the present
calculation, we employ the NL3 parameter set \cite{lal}. To include the
effect of pairing , the occupancies  have been
modified in the framework of BCS approximation \cite{agr}. The
single-particle states are calculated using spherical oscillator basis with
twelve shells. The values of the chemical  potential and the pairing gap at
a given temperature are calculated using all the single-particle states up
to $2\hbar\omega_0$ (the model space) above the Fermi surface without
assuming any core.  With increase of the basis space to twenty shells and the
model space to $3\hbar\omega_0$, the changes in the results
are insignificant. The continuum
corrections are not taken into account in the temperature range explored
($T \le 3$ MeV) where it has earlier \cite{bon,agr1} been found to be quite
small. Once the set of single-particle states are obtained in the RMF
theory, the shell corrections are evaluated in the standard Strutinsky
\cite{str} prescription. 
The shell correction energy $\Delta_{sc}$ is calculated as 
\begin{equation}
\Delta_{sc} = \sum_i n(\epsilon_i,\lambda,T)\epsilon_i - \int
\tilde{g}(\epsilon)\epsilon n(\epsilon,\tilde{\lambda},T) d\epsilon
.
\end{equation}
In eq. (1), the sum runs over all the single-particle states with energy
$\epsilon_i$ in the model space. The function $n$ is the
occupancy and $\tilde{g}(\epsilon)$ is the smoothened density of
single-particle states \cite{rin}. The chemical potentials $\lambda$ and
$\tilde{\lambda}$ are obtained for the discrete and the smoothened
single-particle states, respectively.
Results for $\Delta_{sc}$ calculated in this method taking a fixed set of
temperature independent single particle levels are found to be in
consonance with those obtained by Bohr and Mottelson \cite{boh} using a
different prescription.

In Fig. 1 , in the  top panel, the changing deformation as a function of
temperature for the nuclei $^{148}Sm$ and $^{150}Sm$ are displayed. In the
bottom panel, the thermal evolution of the neutron single-particle level
spectrum of the nucleus $^{150}Sm$ is also shown. The qualitative features
of the evolving  proton single-particle spectrum for $^{150}Sm$ or the
neutron/proton spectra for the nucleus $^{148}Sm$ are similar and  are
therefore not displayed here. At $T = 0$, the preferential occupation of
the $m-$orbitals near the Fermi surface stemming  from the $j-$state
forces the system towards a static ground state deformation. As the system
heats up, the occupancies of different $m-$states evolve in a
self-consistent manner and drive the single-particle potential towards a
more spherically symmetric one which in turn reorganizes the single-particle
orbitals. This reorganization becomes very fast at the shape transition
temperature $T_p$ leading to a sudden collapse 
of the deformation. This is evident from 
both the panels of Fig. 1. 

The thermal evolution of $\Delta_{sc}$ is shown for $^{150}Sm$ in the top
panel of Fig. 2.
This nucleus is deformed in its ground state with $\beta_2 = 0.19$. With
the set of single-particle energy states fixed corresponding to this ground
state, the calculated $\Delta_{sc}$ is shown as the dotted line. 
The dashed-dot line refers to
$\Delta_{sc}$ calculated with  the set of single-particle states
corresponding to the spherical configuration of this nucleus as obtained at
the shape transition temperature $T_p$. It may be mentioned that for a
fixed spherical shape, the single-particle states are practically
independent of temperature below $T\sim 4$ MeV \cite{egi,sau} and therefore
these set of states have been used for the calculation of
$\Delta_{sc}$ (dashed-dot line). In a fully self-consistent calculation, as
expected, $\Delta_{sc}$ (solid line) coincides at $T = 0$ with the dotted
curve and merges with the dashed-dot curve at and beyond the transition
temperature $T_p$. The transition temperature so calculated for this nucleus 
and for other nuclei are in agreement with those calculated recently
also in the RMF framework by Gambhir {\it et al} \cite{gam1}. 
As the temperature approaches $T_p$, there is a sudden
drop in deformation (see Fig. 1), with an abrupt enhancement of $\sim 3$
MeV in $\Delta_{sc}$. Beyond this temperature, the nucleus is spherical and
$\Delta_{sc}$ decreases monotonically. A nucleus in a spherical equilibrium
configuration in its ground state can not have a positive shell correction
energy. Strikingly, here we find that though the shape becomes spherical at 
and beyond
$T_p$, the shell correction energy is positive (more on this is discussed
later). In the bottom panel of Fig. 2, the self-consistent
thermal evolution of $\Delta_{sc}$ for the nuclei $^{148}Sm$, $^{152}Dy$
and $^{154}Dy$ are shown. They exhibit the same kind of structure as in the
case of $^{150}Sm$.

The self-consistent thermal evolution of the shell correction energy is
also studied for two deformed lighter nuclei, namely, $^{64}Zn$ and $^{66}Zn$.
In Fig.3, the temperature evolution of deformation (top panel) and the
shell correction energy $\Delta_{sc}$ (bottom panel) are displayed for
these two systems. It is found that the collapse of the deformation
is a little smoother for these two nuclei compared to the isotopes
of $Sm$ and $Dy$ studied. A broad but less prominent bump in
$\Delta_{sc}$ around the transition temperature is also seen. 
As pointed out earlier, the thermal evolution of $\Delta_{sc}$ 
depends on two competing
mechanisms, i.e., the smoothening of the fluctuation effects of
the single-particle spectra around the Fermi surface and the bunching 
of different $m$-states with increasing temperature. The counter-balance 
of these two effects determines the details of the structure in the
temperature dependence of $\Delta_{sc}$ for deformed nuclei.
This is reflected in Fig.2 as well as in Fig.3.

Since  spherical nuclei in ground state have negative shell
corrections, the positive shell correction energy for the systems turned
spherical at or beyond the shape transition temperature at first glance may
look counter-intuitive. This, however, can be understood from an
examination of the level density of the single-particle states in the
vicinity of the Fermi surface within the smearing width taken to be $\sim
1.2\hbar\omega_0$. We are dealing with open-shell deformed nuclei with
temperature induced spherical shapes that have different single-particle
structure near the Fermi surface as compared to those in the regular spherical
(i.e. closed-shell) nuclei.  Thus, one
need  not expect the same behaviour of the shell corrections though the
systems are spherical in both cases. In Fig. 4, the results for the shell
correction energy $\Delta_{sc}$ for neutrons  and protons are given
separately for systems $^{150}Sm$ (top panel) and $^{64}Zn$ (bottom panel).
It is found that $\Delta_{sc}$ remains positive for protons and negative
for neutrons in case of $^{150}Sm$ and the situation is reversed for
$^{64}Zn$. The number of neutrons in $^{150}Sm$ is closer to a magic number
(N=82) whereas the number of protons correspond to a mid-shell. This
explains the sign of the shell corrections  for neutrons and protons; the
total shell correction is the sum of these two. For $^{64}Zn$,
the sign of the proton and neutron shell correction is reversed; 
here the proton number as compared to the number of neutrons is closer
to the magic number 28.

That the shell correction energy dissolves with increasing temperature for
a fixed set of single-particle levels is well-known.
For deformed nuclei, because of the self-consistent reorganization of the
single-particle field, we now find that there is a  strong   
build-up of the shell
correction energy at the shape transition temperature. This is a
manifestation of  the delicate interplay between the self-consistent
evolution of the single-particle states with temperature and their
occupancies. In phenomenological analyses of the 
temperature dependence of the level density of deformed
nuclei, the temperature induced shell correction would  have a role
to play. The temperature dependent shell effects are generally included
using a single wave fluctuation of the level density \cite{ram,shl}
that corresponds to a fixed set of single particle states; as is seen,
for a proper evaluation of the level density for deformed systems
the reorganisation of the single particle spectra with temperature
need to be considered.

The authors gratefully acknowledge helpful discussions with Prof. S. S.
Kapoor.

\newpage
\noindent {\bf Figure Captions}
\begin{itemize}
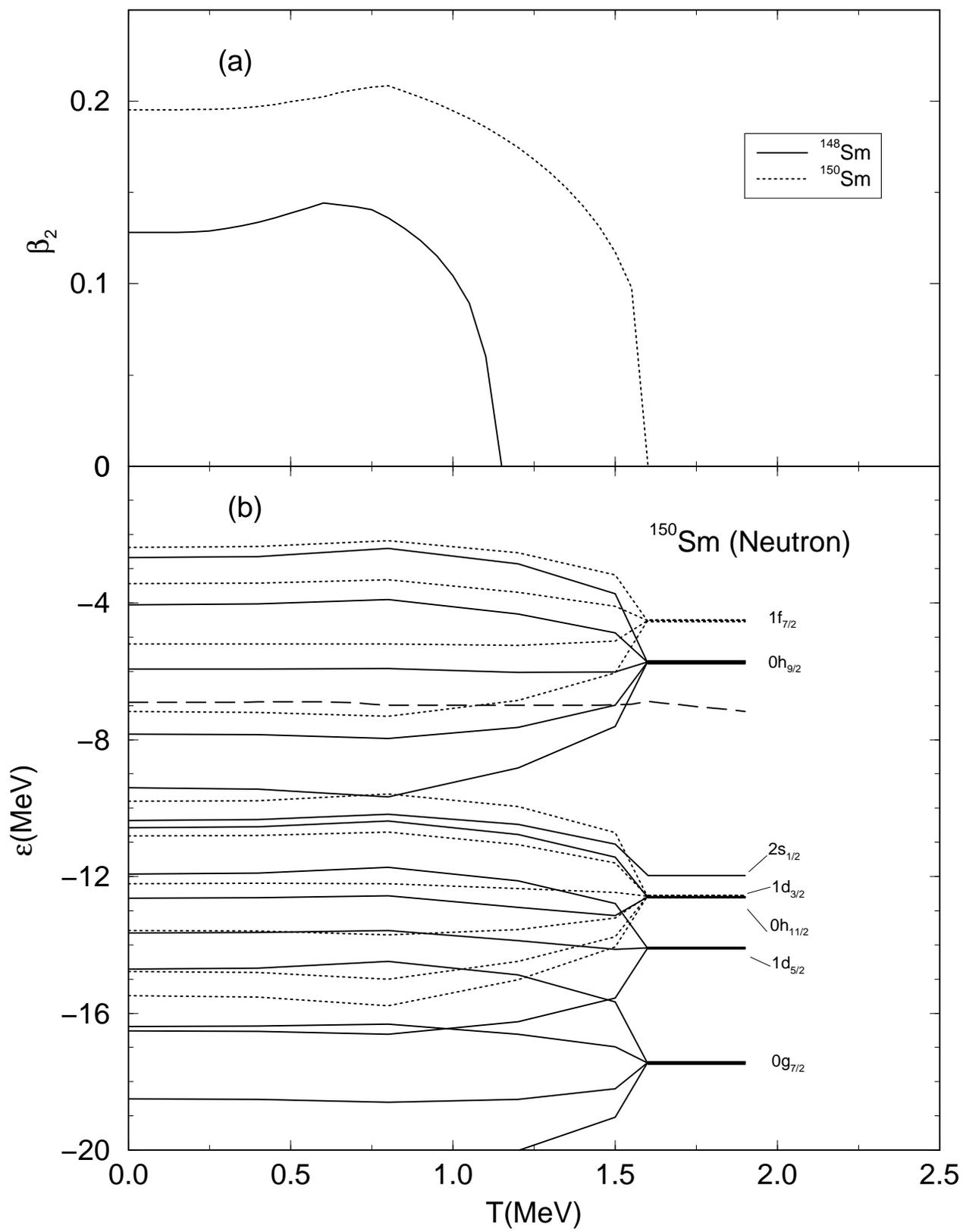
\item[Fig. 1] (a) The evolution of the deformation $\beta_2$ as a function
of temperature  for $^{148}Sm$ and $^{150}Sm$. (b)  The thermally evolving
single-particle neutron spectrum for $^{150}Sm$.
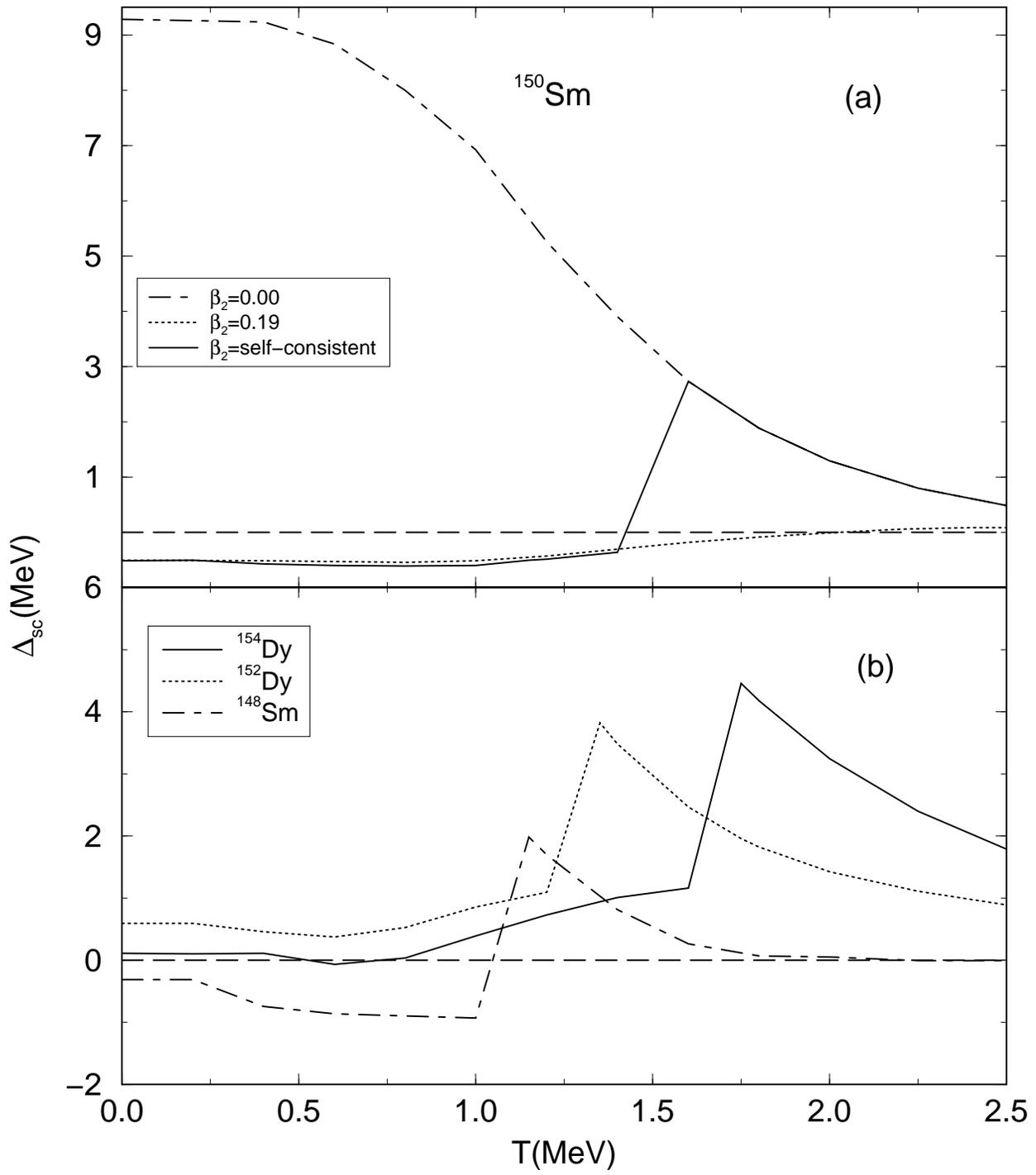
\item[Fig. 2] (a) The shell-correction energy for $^{150}Sm$ is shown for
different choices of single-particle states (see text). (b) The
self-consistent evolution of $\Delta_{sc}$ as a function of temperature for
the nuclei $^{148}Sm$, $^{152}Dy$ and $^{154}Dy$.
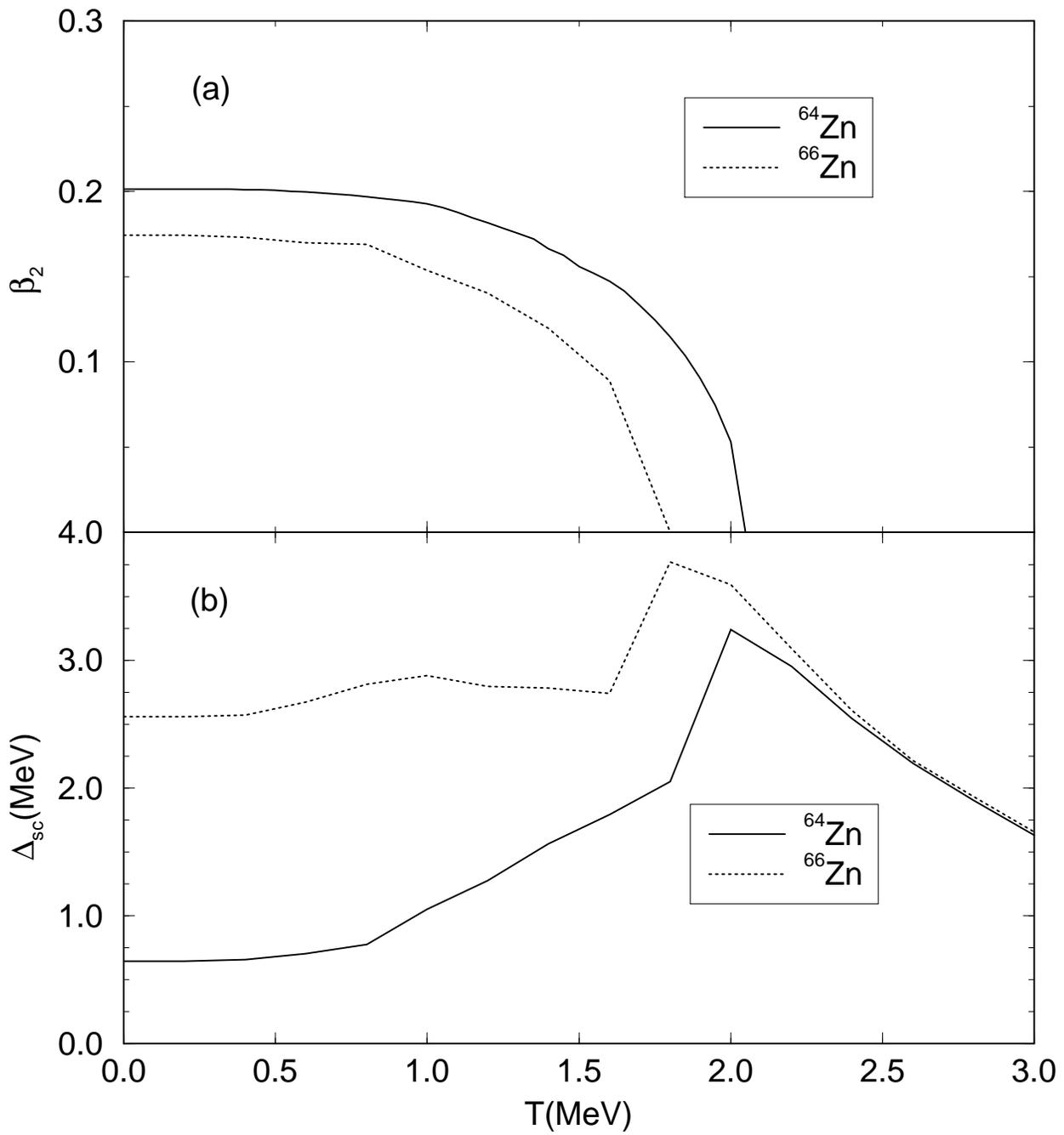
\item[Fig. 3] (a) Same as Fig. 2a for the nuclei $^{64}Zn$ and $^{66}Zn$.
(b) Same as Fig. 3b for the systems $^{64}Zn$ and $^{66}Zn$.
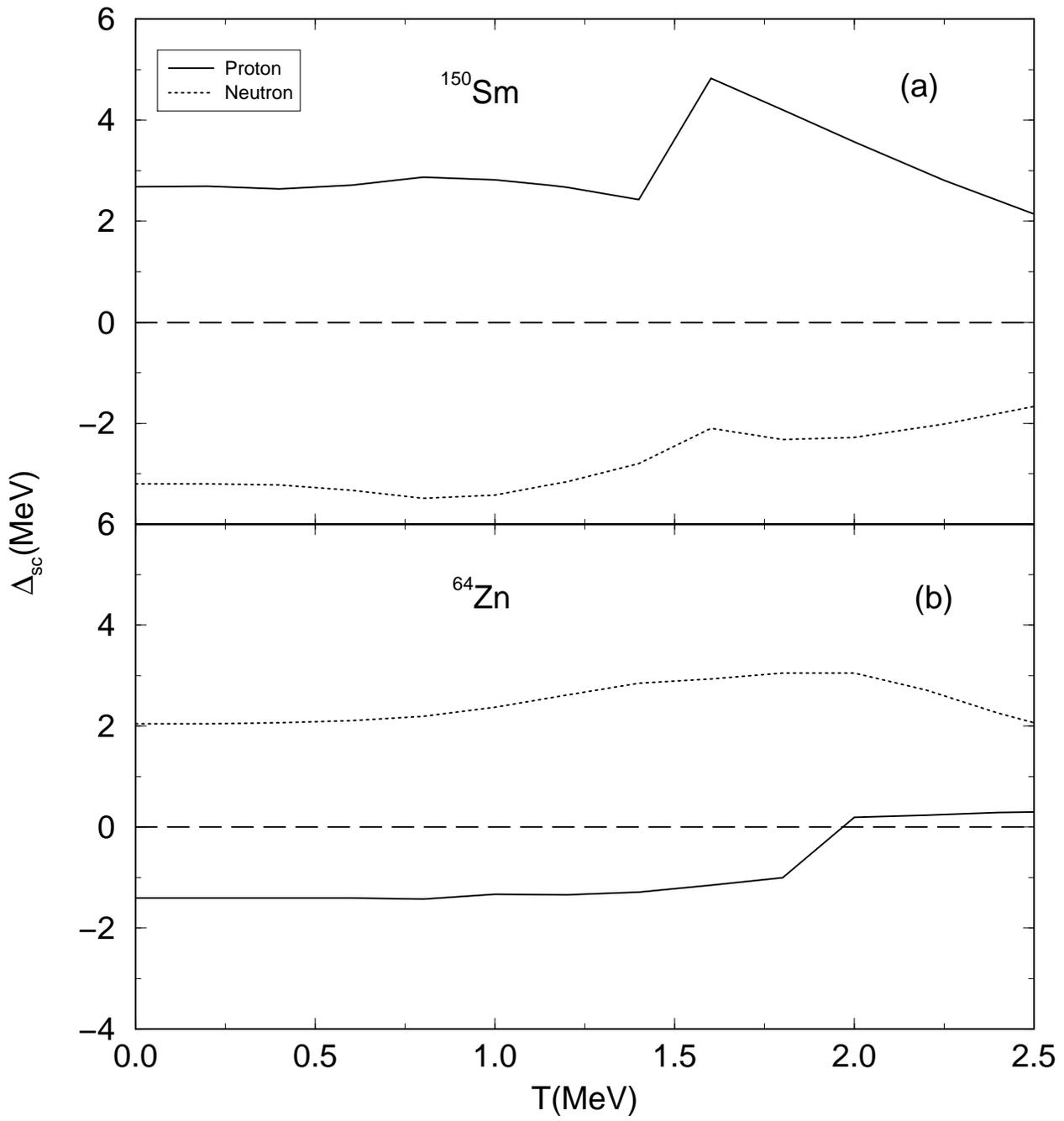
\item[Fig. 4] The neutron and proton contributions to the total
shell-correction energy for the systems $^{150}Sm$ and $^{64}Zn$.
\end{itemize}

\newpage
\begin{thebibliography}{99}
\bibitem{str} V. M. Strutinsky, Nucl. Phys. {\bf A95}, 420 (1967).
\bibitem{ram} V. S. Ramamurthy, S. S. Kapoor and  S. K. Kataria , Phys.
Rev. Lett. {\bf 25}, 386 (1970).
\bibitem{bra} M. Brack and P. Quentin, Phys. Scr. {\bf A10}, 163 (1974).
\bibitem{goo} A. L. Goodman, Phys. Rev. {\bf C33}, 2212 (1986).
\bibitem{egi} J. L. Egido, L. M. Robledo and V. Martin, Phys. Rev. Lett. 
{\bf 85}, 26 (2000).
\bibitem{gam} Y. K. Gambhir, P. Ring and A. Thimet, Ann. Phys. (N.Y.),
{\bf 198}, 132 (1990).
\bibitem{agr} B. K. Agrawal, Tapas Sil, S. K. Samaddar and J. N. De, Phys.
Rev. {\bf C63}, 024002 (2001).
\bibitem{lal} G. A. Lalazissis, J. K$\ddot{o}$nig and P. Ring, Phys. Rev.
{\bf C55}, 540 (1997).
\bibitem{bon} P. Bonche, S. Levit and D. Vautherin, Nucl. Phys.{\bf
A436}, 265 (1985).
\bibitem{agr1} B. K. Agrawal, Tapas Sil, J. N. De and S. K. Samaddar, 
     Phys. Rev. {\bf C62}, 044307 (2000).
\bibitem{rin} P. Ring and P. Schuck, The Nuclear Many-Body Problem, p. 89
(Springer, Berlin, 1980).
\bibitem{boh} A. Bohr and B. Mottelson, Nuclear Structure, Vol. II,  p. 609, 
    W. A.  Benjamin, Inc.,  (1975).
\bibitem{sau} G. Sauer, H. Chandra and U. Mosel, Nucl. Phys. {\bf A269},
221 (1976).
\bibitem{gam1} Y. K. Gambhir, J. P. Maharana, G. A. Lalazissis, C. P.
Panos and P. Ring, Phys. Rev. {\bf C62}, 054610 (2000).
\bibitem{shl} S. Shlomo and J. B. Natowitz, Phys. Lett. {\bf B252}, 
187 (1990).
\end{thebibliography}
\end{document}